\documentclass[aps,prl,showpacs,superscriptaddress,twocolumn]{revtex4-2}
\usepackage[utf8]{inputenc}
\usepackage{amsmath}
\usepackage{amssymb}
\usepackage{color}
\usepackage{verbatim}
\usepackage{array}
\usepackage{graphicx}
\usepackage{bm}
\usepackage[ragged]{sidecap}
\usepackage{dsfont}
\usepackage{ulem}
\usepackage{mathrsfs}

\usepackage{gensymb}
\usepackage{siunitx}
\usepackage{url}
\usepackage{dcolumn}
\usepackage{soul}
\newcolumntype{d}{D{.}{.}{-1}}
\newcolumntype{f}[1]{D{.}{.}{#1}}

\usepackage{hyperref}
\hypersetup{
    colorlinks,
    citecolor=blue,
    filecolor=orange,
    linkcolor=cyan,
    urlcolor=green}

\newcommand{\eg}{{\textit{e.g., }}}
\newcommand{\ie}{{\textit{i.e., }}}

\begin{document}

%========= TITLE =====================================================

\title{Testing Quantum Electrodynamics with Exotic Atoms}

\author{Nancy Paul} 
\email{npaul@lkb.upmc.fr}
\affiliation{Laboratoire Kastler Brossel, Sorbonne Universit\'e, CNRS, ENS-PSL Research University, Coll\`ege de France, Case\ 74;\ 4, place Jussieu, F-75005 Paris, France}

\author{ Guojie Bian}  
\email{bianguojie@gmail.com}
\affiliation{Laboratoire Kastler Brossel, Sorbonne Universit\'e, CNRS, ENS-PSL Research University, Coll\`ege de France, Case\ 74;\ 4, place Jussieu, F-75005 Paris, France}
\affiliation{Key Laboratory of Computational Physics, Institute of Applied Physics and Computational Mathematics, 100088 Beijing, China}
\author{Toshiyuki Azuma}  
\email{toshiyuki-azuma@riken.jp}
\affiliation{RIKEN, Wako, Saitama 351-0198, Japan}

\author{Shinji Okada }  
\email{sokada@isc.chubu.ac.jp}
\affiliation{Chubu University, Kasugai, Aichi 487-8501, Japan.}

\author{ Paul Indelicato} 
\email[Corresponding Author: ]{paul.indelicato@lkb.upmc.fr}
\affiliation{Laboratoire Kastler Brossel, Sorbonne Universit\'e, CNRS, ENS-PSL Research University, Coll\`ege de France, Case\ 74;\ 4, place Jussieu, F-75005 Paris, France}
\homepage{http://www.lkb.upmc.fr/metrologysimplesystems/project/paul-indelicato/}

%========= DATE ======================================================

% It is always \today, today,
\date{\today}

%========= ABSTRACT ==================================================

%for EPJD: 
%\abstract{
%Li-like Ar
%}

\begin{abstract}
Precision study of few-electron, high-$Z$ ions is a privileged field for probing high-field, bound-state quantum electrodynamics (BSQED). However, the accuracy of such tests is plagued by nuclear uncertainties, which are often larger than the BSQED effects under investigation. We propose an alternative method with exotic atoms, and show that transitions may be found between circular Rydberg states where nuclear contributions are vanishing while BSQED effects remain large. When probed with newly available quantum sensing detectors, these systems offer gains in sensitivity of  \numrange{1}{2} orders of magnitude, while the mean electric field in these systems largely exceeds the Schwinger limit.
\end{abstract}

%========= CLASSIFICATION ============================================
%========= to be checked ============================================
%
%\pacs{36.10.Ee,31.30.jr,36.10.Gv}
\maketitle

Quantum electrodynamics is the best understood quantum field theory, and serves as the foundation for searches for Beyond Standard Model physics with atoms at the precision frontier (see \cite{sbdk2018} for a recent review). Bound-state QED (BSQED) allows to perform extremely precise calculations for few-electron systems, where for example, a full evaluation of BSQED effects can be achieved in hydrogen with \num{12} significant figures \cite{mnt2016}.
However the evaluation of BSQED contributions is not amenable to a converging $Z\alpha$ expansion, as the series is only asymptotic \cite{moh1974}, even at moderate values of $Z$.  The best tests of strong-field BSQED may come then from few-electron, highly-charged ions (HCI). Measurements have been performed in hydrogen-like ions up to U$^{91+}$ and in other simple systems (see, \eg \cite{ind2019} for a recent review). Few parts-per-million (ppm) accuracy has been achieved for medium-$Z$ atoms with electron cyclotron ion sources or electron beam ion traps, and crystal spectrometers \cite{bbkc2007,kmmu2014,esbr2015, asgl2012,mssa2018,mbpt2020,assg2014}. The highest-$Z$ species are studied at accelerators and storage ring facilities, like the GSI/FAIR facility \cite{bifl1991,lbfb1994,bmlg1995,gsbb2005,gths2018,laab2016} with the aim of improving accuracy from \SI{50}{ppm} down to \SI{10}{ppm}.

Despite the success of BSQED, its completeness has been called into question in light of numerous recent experimental results. Since \num{2010}, there exists a severe discrepancy between the proton charge radius as extracted from normal versus muonic hydrogen, dubbed the \textit{proton size puzzle} \cite{pana2010,ansa2013,pnfa2016}. While recent experimental results have reduced this discrepancy \cite{bmmp2017,bvhm2019}, the problem persists \cite{fgtb2018} and has spurred numerous propositions for new physics beyond the standard model (see, \eg \cite{cnw2014,kmp2014,aaaa2015,btad2015,caf2015,lam2015,lmm2016,dffs2017,mfkn2018,sbdk2018,sta2018}). The recent result in positronium (Ps), a pure QED system, shows a \num{4.5}$\sigma$ discrepancy between experiment and theory \cite{gbhc2020}, compounding a more than \num{20} years-old \num{3}$\sigma$ discrepancy on the Ps ground state hyperfine structure \cite{hwrc2018,pak1998}. \\
%%%%%%%%%% Fig. 1 %%%%%%%%
\begin{figure}[tp]
\centering
        \includegraphics[width=\columnwidth]{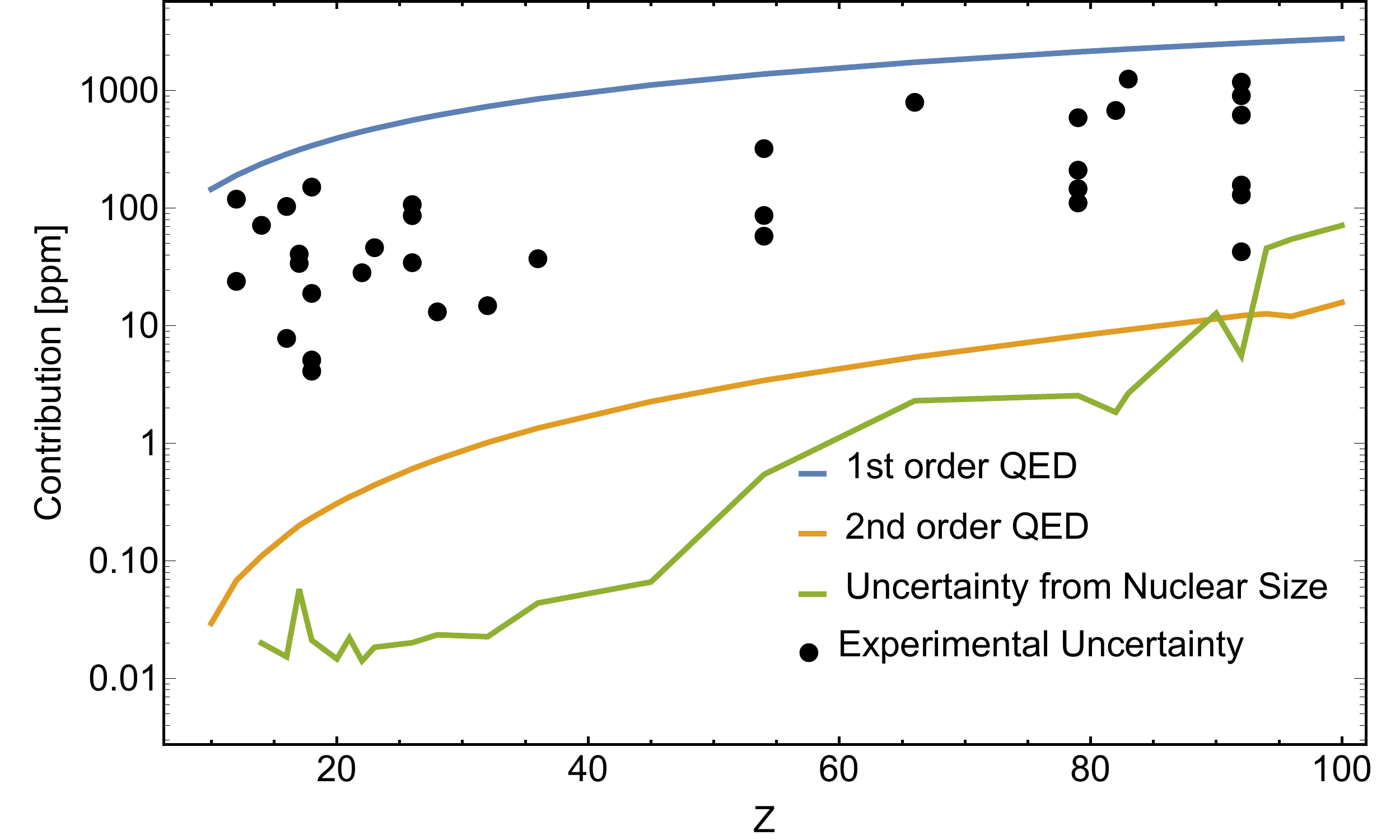}
\caption{Size of  the QED contributions to the Ly$\alpha$ transition energy, scaled by $Z^4$ as a function of $Z$, and comparison with experimental uncertainty and uncertainty on the theoretical energy due to uncertainties in the nuclear size \cite{aam2013}. Refs. for experiments: \cite{astf1980,bciz1990,bdfl1985,bfli1993,bifl1991,biss1989,blbf1994,bmic1983,bmlg1995,btim1983,cldh2007,clsd2009,dsj1985,gcph2010,gsbb2005,gths2018,hfkb1998,kabe2017,kkrs1984,kmmu2014,ldhs1994,mrkk1986,rsdc1984,sbbt1982,smbb1993,smbd2000,smgm1992,smlr1987,tas2002,tbil1985,tgbb2009,tkbb2002,wbbl2000}. Theory from \cite{yas2015}.
\label{fig:h-like}
}
\end{figure}
%%%%%%%%%%%%%%
%
For medium and high-$Z$ ions, where BSQED effects are largest, attaining for example \SI{256}{\electronvolt} for the Ly$\alpha$ lines in U$^{91+}$, there remains a clear lack of accurate measurements (around or below \SI{10}{ppm}) that would permit detailed tests of the theory \cite{ind2019}. This is because few-electron, high-$Z$ ions must be created at accelerator facilities, and large Doppler effects from in-flight spectroscopy presently dominate the uncertainties. This may be seen in Fig. \ref{fig:h-like} which shows the experimental uncertainties and theoretical contributions and uncertainties for the $2p-1s$ transitions in one-electron atoms used for probing high-field BSQED. More precise measurements will not give the full answer, because uncertainties due to finite nuclear size effects (FNS) and nuclear polarization or deformation cannot be evaluated with sufficient accuracy for high-$Z$ ions. The  uncertainty on these nuclear effects in atomic transition energies represent a sizable fraction of the second-order BSQED \cite{yis2003} as shown in Fig.~\ref{fig:h-like}. 

In this Letter, we propose an alternative method for probing high-field BSQED using transitions between circular Rydberg states in exotic atoms. Up to now, exotic atoms, where an electron is replaced by another particle, \eg a muon ($\mu^{-}$), a pion, or an antiproton ($\bar p$), have been mostly used to measure nuclear properties, strong interaction effects or particle mass. As these particles are much heavier than the electron, they are closer to the nucleus (\eg \num{207} times for the $\mu^-$) and thus very sensitive to the nuclear density distribution. The nuclear radii and shape of most stable elements have been obtained by combining muonic atom spectroscopy, electron scattering and optical or x-ray measurements of the isotopic shift (see, \eg \cite{rkkm1977,sspp1982,fbhs1995,aabc1999,gaab1999}). However, in these systems, the proximity of the exotic particle to the nucleus also enhances the BSQED contributions to the transition energies, as the particle experiences higher electric fields with respect to normal atoms. As will be shown in the following, a region of transitions may be found between high-$n$ circular Rydberg states where nuclear contributions to the transition energies are vanishing, while BSQED contributions remain large, creating a unique opportunity to cleanly probe high-field BSQED.

Precision studies of exotic atoms are now possible due to the development of a new generation of microcalorimeter detectors, with improved resolution and accuracy, tested successfully in the harsh conditions of high-energy accelerators \cite{obdf2014,obcd2016,kabe2017,oabc2020}. An accuracy of \SI{0.04}{\electronvolt} has been attained for \SI{6.4}{\kilo\electronvolt} \cite{tdbc2016}, and resolutions of $\Delta E/E \sim 10^{-4}$\cite{oabc2020,fabd2017,tdbc2016}.  In the  \SIrange{50}{100}{\kilo\electronvolt} range, a resolution of  \SI{22}{\electronvolt} \cite{tdbc2016} at \SI{97.4}{\kilo\electronvolt} has been achieved, while an accuracy  of  \SI{0.4}{\electronvolt} \cite{kbcv2020} was demonstrated for the $^{241}$Am $\gamma$-ray line at  \SI{60}{\kilo\electronvolt}. Detection ranges can be pushed up to \SI{200}- \SI{400}{\kilo\electronvolt}  \cite{bhsh2012,nmbb2013}. The improvement in resolution compared to conventional solid-state detectors allows to disentangle the fine structure of the transitions and thus an improved comparison with theory. Microcalorimeter detectors also boast quantum efficiencies of up to $\sim$ 70\% , and total collection efficiencies of \num{3E-3} around \SI{100}{\kilo\electronvolt} and still \num{7E-4} at \SI{200}{\kilo\electronvolt} \cite{bhsh2012}. This is to be compared with transmission crystal spectrometers that achieve resolutions of $\Delta E/E \sim 8\times 10^{-4}$, but collection efficiencies of only around \num{8E-8} \cite{bgth2015}. 

We have performed calculations using the Multiconfiguration Dirac-Fock General Matrix Elements (MCDFGME) code, which can evaluate energies, transition probabilities, and hyperfine structure for exotic atoms composed of an arbitrary number of electrons and of a fermion or a boson \cite{mdf1978,spbi2005,tai2007,grfi2008,ind2013}, even for very high-$n$, high-$\ell$ states.
The energies of highly-charged, muonic, and antiprotonic atoms are obtained using a full atomic wavefunction, composed of a determinant with all the electrons, multiplied by the exotic particle wavefunction, and solving the full coupled system of differential equations. The electron-electron interaction as well as the particle-electron interaction is chosen to be the full Breit operator with Coulomb, magnetic and retardation in the Coulomb gauge. The influence of the exotic particle on electron BSQED corrections and vice-versa is also taken into account. For the antiproton, its finite size and its $g$-factor are also accounted for.  The Uehling contribution to the vacuum polarization is  evaluated to all orders \cite{ind2013} thus allowing to include the loop-after-loop  second-order corrections in the case of exotic atoms.  The Wichmann and Kroll potential of order $\alpha (Z\alpha)^3$, $\alpha (Z\alpha)^5$ and $\alpha (Z\alpha)^7$ \cite{wak1956,blo1972,far1976}, as well as the K\`all\"en and Sabry \cite{kas1955,far1976} $\alpha^2 (Z\alpha)$ second order potential are also included.
 %[164], 
The finite nuclear size correction (FNS) is treated with the Fermi model using nuclear radii from \cite{aam2013} with a thickness parameter of \SI{2.3}{fm}. Nuclear deformation and polarization effects are not included as they are negligible for the high-$n$ circular Rydberg states we consider here.

\begin{figure}[htbp]
\includegraphics[width=\columnwidth]{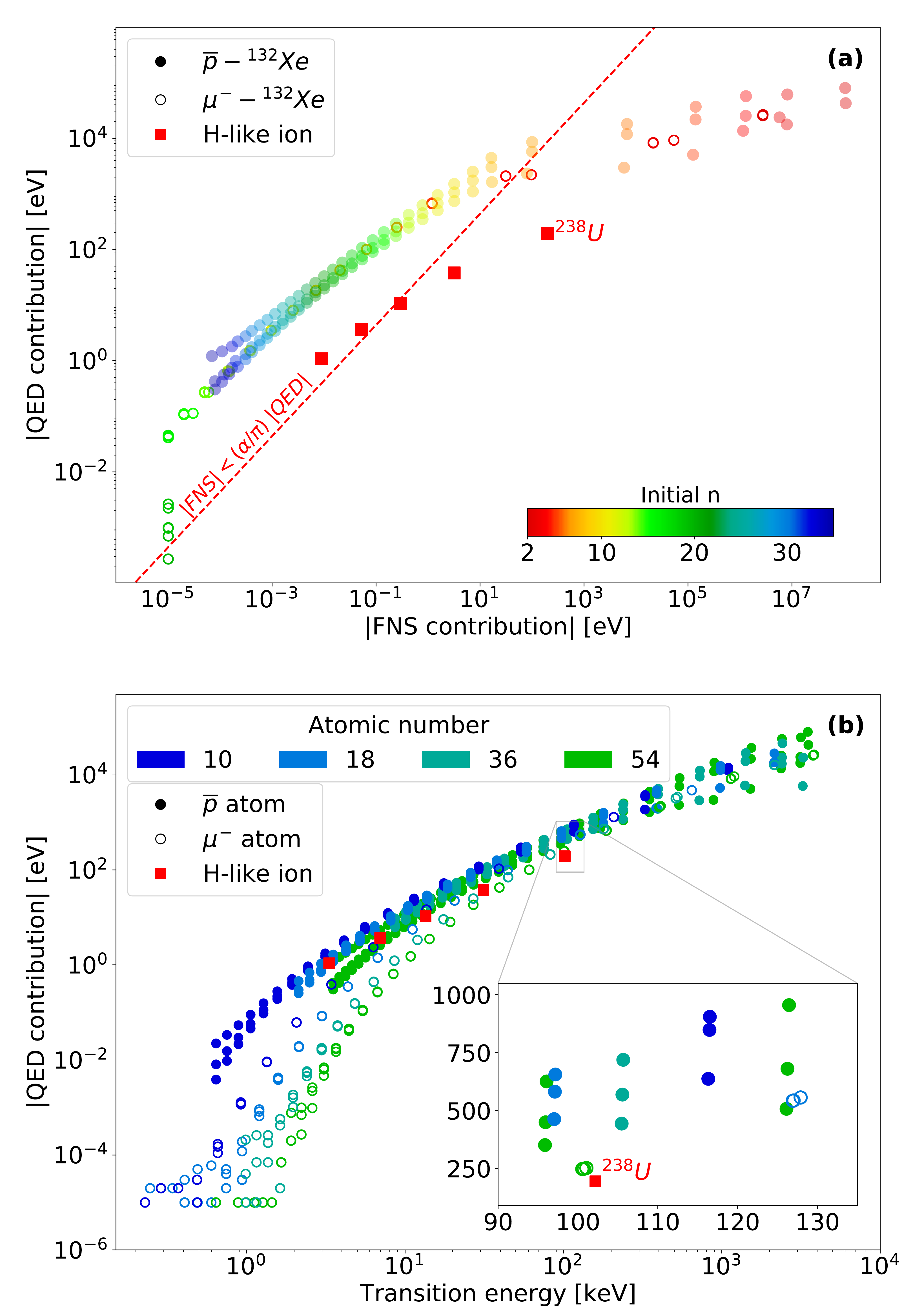}
\caption{(a) QED contribution to the transition energy as a function of the finite nuclear size (FNS) contribution to the transition energy for yrast transitions in antiprotonic and muonic $^{132}$Xe.
(b) QED contribution to the transition energy as a function of the transition energy for yrast transitions in antiprotonic and muonic noble gases. In both (a) and (b), the $2p-1s$ transitions in H-like ions ($Z=18,26,36,54,92$) are shown for reference.}
\label{fig:QEDvsFNS}
\end{figure}

MCDFGME calculations were performed for levels starting from $n=15$ for Ne up to $n=35$ to U, for all possible $\ell$ and $j$ quantum numbers as the exotic particles capture onto high-$n$ orbitals according to $n_{exotic}\sim n_{e^{-}} \sqrt{m_{exotic}/m_{e^-}}$ \cite{bak1960}. 
Assuming a statistical $(2\ell+1)$ population of the state in which the exotic particle is captured, and absorption of low angular momentum states by the nucleus, after a phase of Auger electron emission, the radiative transitions preferentially populate circular Rydberg states, $(n, \ell=n-1)$ \cite{grfi2008}. We have thus studied the relative influence of the BSQED and FNS corrections to the yrast cascade transitions. Results for the yrast states in $\mu$Xe and $\bar p$Xe   are shown in Fig.~\ref{fig:QEDvsFNS}~(a). It shows that BSQED contributions are significantly enhanced with respect to the FNS contribution, by one or two orders of magnitude compared to  $2p-1s$ transitions in HCI, also shown in Fig.~\ref{fig:QEDvsFNS}~(a). The best cases for tests of higher-order BSQED corrections are found when second-order BSQED corrections are much larger than the contribution from the nuclear size, \ie larger than the effect of nuclear size uncertainty. This limit is shown as the red dashed line in Fig.~\ref{fig:QEDvsFNS}~(a). The exotic atom transitions falling below this line are either high-$n$ states far from the nucleus, where QED effects are also small, or transitions between low-$n$ orbitals close to the nucleus, where the FNS effects become too large.

For a given energy range, transitions in exotic atoms also have larger BSQED contributions than their HCI equivalents. This may be seen in Fig.~\ref{fig:QEDvsFNS}~(b), which shows the magnitude of the QED contribution to the transition energy as a function of the transition energy between Rydberg states in muonic and antiprotonic noble gases, and $2p-1s$ transitions in HCI. 
The first observation is that for a given transition energy, antiprotonic atoms always have the largest QED contributions, as the antiproton is heavier than both the muon and the electron and thus is closer to the nucleus and feels a higher Coulomb field. Secondly, with exotic atoms, large QED contributions can be accessed with much lower-$Z$ systems. This can be seen in the inset of Fig.~\ref{fig:QEDvsFNS}~(b), which highlights the region around \SI{100}{\kilo\electronvolt}, generally accessible with Ge detectors and where one finds the $2p-1s$ transition in H-like U, considered today to be the best test of high-field BSQED \cite{gsbb2005}. In this  energy range, exotic atoms have BSQED contributions enhanced by factors of \numrange{3}{5} and for atoms of much lower-$Z$, thus being easier to access in the laboratory. 

A detailed comparison of the transitions used for QED tests in HCI and proposed transitions for QED tests with exotic atoms is shown in Table~\ref{tab:energies}. For all cases shown here, the BSQED contributions are much larger and the FNS corrections much smaller in exotic atoms than they are in the Ly$\alpha$ lines of similar energies, showing quantitatively the general trends presented in Fig.~\ref{fig:QEDvsFNS}~(a). Second-order QED effects notably start to become measurable, even for low-$Z$ systems, and for example in antiprotonic atoms are \numrange{1}{2} orders of magnitude larger than for the HCI Ly$\alpha$ transitions in the same energy range. 

One complication when using exotic atoms for high-precision measurements is the presence of electrons not ionized during the cascade. For example, the uncertainty in the number of remaining electrons in pionic Mg due to the use of a solid target and electron recapture led to a problem in the determination of the  pion mass from crystal  spectroscopy \cite{jgl1994}. The use of low-pressure gas targets avoids recapture \cite{lbgg1998,tabd2016}, and measurements on high-$Z$ elements beyond Xe can be achieved with gaseous organometallic compounds like WF$_6$ or UF$_6$. These molecules undergoes a \textit{Coulomb explosion} during the exotic particle capture, leading to a negligible broadening of the emitted x-rays as observed in other molecular gases \cite{sabg2000}. Study of the cascade in antiprotonic rare gases has shown that there are no remaining electrons in $\bar p$Ne and $\bar p$Ar \cite{bbgh1988}. Nevertheless, as the efficiency of the Auger mechanism varies with the mass of the exotic particle and the $Z$ of the atom, there may be an uncertainty in the number of remaining electrons for heavier elements, which can partly be removed by a study of the cascade with higher-resolution detectors.

The effect of remaining electrons can be calculated quite accurately with the MCDFGME code \cite{mdf1978,vad1978,grfi2008}. Results for \num{2} and \num{10} remaining electrons are shown in Table~\ref{tab:elec_screen}.  It shows that the screening effect is strongly dominated by the electrons in the  $1s$ orbital, and that the uncertainty due to this effect can be minimized using low-  to medium-$Z$ antiprotonic atoms, where the Auger mechanism is most efficient, and low-$n$ states where the screening shifts are smallest. It also shows that an exact knowledge of the number of electrons is not needed because, \eg the shift does not change significantly going from \numrange{2}{10} electrons.

Another issue specific to antiprotonic atoms is the possible hadronic shift due to the strong interaction between the nucleus and the antiproton. It  has been shown that  the highest circular states affected by annihilation are $\bar p$Ne, $\bar p$Ar, $\bar p$Kr and $\bar p$Xe are $(n_{\bar p} , \ell_{\bar p}) = (4, 3)$, (5, 4), (7, 6), and (8,7), respectively \cite{pot1979}, and thus we have restricted ourselves to transitions above these $(n_{\bar p} , \ell_{\bar p})$ values. 

%%%%%%%%%%%

%quantum sensing detectors
%realistic experimental prospects, JPARC, CERN-ELENA
\begingroup
\squeezetable
\begin{table*}							
\caption{Comparison of transition energies, QED and finite size corrections for transitions amenable to BSQED tests in muonic atoms and antiprotonic atoms, compared to transitions of similar energy in highly-charged ions. All energies are given in \si{\electronvolt}. 
}
\label{tab:energies} 							
\begin{ruledtabular}
\begin{tabular}{ccccdddddrrr}	
\multicolumn{1}{c}{Particle}	&	\multicolumn{1}{c}{Element}	&	\multicolumn{1}{c}{Initial}	&	\multicolumn{1}{c}{Final}	&	\multicolumn{1}{c}{Theo. Trans.}	&	\multicolumn{1}{c}{1st order}	&	\multicolumn{1}{c}{2nd order}	&	\multicolumn{1}{c}{$g-2$ $\bar{p}$}	&	\multicolumn{1}{c}{FNS}	&	\multicolumn{1}{c}{FNS/QED}	&	\multicolumn{1}{c}{Exp.}	&	\multicolumn{1}{c}{Ref.}	\\
	&		&	\multicolumn{1}{c}{state}	&	\multicolumn{1}{c}{state}	&	\multicolumn{1}{c}{energy}	&	\multicolumn{1}{c}{QED}	&	\multicolumn{1}{c}{QED}	&		&		&		&		&		\\
	\hline																							
	$e^{-}$	&	$^{40}$Ar	&	$2p_{3/2}$	&	$1s_{1/2}$	&	3322.9931	&	-1.1238	&	0.0007	&		&	-0.0090	&	\SI{0.804}{\percent}	&	\num{3322.993+-0.014}	&	\cite{kmmu2014}	\\
	$\mu^{-}$	&	$^{20}$Ne	&	$6h_{11/2}$  	&	$5g_{9/2}$  	&	3419.6826	&	0.3844	&	0.0042	&		&	0.0001	&	\SI{0.013}{\percent}	&		&		\\
 	$\bar{p}$ 	&	$^{40}$Ar	&	$17v_{31/2}$	&	$16u_{29/2}$	&	3522.9849	&	1.2209	&	0.0124	&	0.0618	&	0.0002	&	\SI{0.014}{\percent}	&		&		\\
 	\hline																							
 	$e^{-}$	&	$^{56}$Fe	&	$2p_{3/2}$	&	$1s_{1/2}$	&	6973.1815	&	-3.8873	&	0.0042	&		&	-0.0527	&	\SI{1.357}{\percent}	&	\num{6972.73+-0.24}	&	\cite{cldh2007}	\\
	$\mu^{-}$	&	$^{20}$Ne	&	$5g_{9/2}$  	&	$4f_{7/2}$ 	&	6297.2613	&	2.3365	&	0.0226	&		&	0.0003	&	\SI{0.013}{\percent}	&		&		\\
	\hline																							
	$e^{-}$	&	$^{84}$Kr	&	$2p_{3/2}$	&	$1s_{1/2}$	&	13508.9648	&	-11.4244	&	0.0181	&		&	-0.2963	&	\SI{2.594}{\percent}	&	\num{13508.95+-0.50}	&	\cite{tbil1985}	\\
	$\mu^{-}$	&	$^{20}$Ne	&	$4f_{7/2}$	&	$3d_{5/2}$  	&	13615.9995	&	14.4762	&	0.1300	&		&	0.0034	&	\SI{0.023}{\percent}	&		&		\\
	$\bar{p}$ 	&	$^{40}$Ar	&	$11n_{21/2}$	&	$10m_{19/2}$	&	13729.2330	&	24.6027	&	0.2198	&	0.9635	&	0.0080	&	\SI{0.032}{\percent}	&		&		\\
	\hline																							
	$e^{-}$	&	$^{132}$Xe	&	$2p_{3/2}$	&	$1s_{1/2}$	&	31283.9469	&	-43.0973	&	0.1068	&		&	-3.1806	&	\SI{7.380}{\percent}	&	\num{31284.9+-1.8}	&	\cite{tgbb2009}	\\
	$\bar{p}$ 	&	$^{20}$Ne	&	$6h_{11/2}$  	&	$5g_{9/2}$  	&	29183.5464	&	106.3706	&	0.9785	&	4.7013	&	0.0719	&	\SI{0.068}{\percent}	&		&		\\
	$\bar{p}$ 	&	$^{84}$Kr	&	$13q_{25/2}$	&	$12o_{23/2}$	&	32965.1441	&	85.8957	&	0.7722	&	5.4910	&	0.0607	&	\SI{0.071}{\percent}	&		&		\\
	\hline																							
	$e^{-}$	&	$^{238}$U	&	$2p_{3/2}$	&	$1s_{1/2}$	&	102175.0991	&	-257.2281	&	1.2278	&		&	-198.5110	&	\SI{77.173}{\percent}	&	\num{102178.1+-4.3}	&	\cite{gsbb2005}	\\
 	$\mu^{-}$	&	$^{132}$Xe	&	$6h_{11/2}$	&	$5g_{9/2}$	&	100690.8146	&	246.4935	&	2.2406	&		&	0.2473	&	\SI{0.100}{\percent}	&		&		\\
	$\bar{p}$ 	&	$^{40}$Ar	&	$6h_{11/2}$	&	$5g_{9/2}$	&	97106.9649	&	524.4499	&	5.1711	&	52.0324	&	1.0708	&	\SI{0.204}{\percent}	&		&		\\
	$\bar{p}$ 	&	$^{84}$Kr	&	$9l_{17/2}$ 	&	$8k_{15/2}$	&	105534.2121	&	505.7735	&	4.8592	&	57.9787	&	1.0625	&	\SI{0.210}{\percent}	&		&		\\
	$\bar{p}$ 	&	$^{132}$Xe	&	$12o_{23/2}$	&	$11n_{21/2}$	&	95937.8536	&	398.8736	&	3.7548	&	46.7550	&	0.7860	&	\SI{0.197}{\percent}	&		&		\\
	\hline																							
	$\mu^{-}$	&	$^{84}$Kr	&	$4f_{7/2}$	&	$3d_{5/2}$  	&	178245.9709	&	712.8889	&	6.7471	&		&	0.4179	&	\SI{0.059}{\percent}	&		&		\\
	$\mu^{-}$	&	$^{132}$Xe	&	$5g_{9/2}$  	&	$4f_{7/2}$ 	&	185826.7311	&	664.4253	&	6.2157	&		&	1.1906	&	\SI{0.179}{\percent}	&		&		\\
	$\bar{p}$ 	&	$^{84}$Kr	&	$8k_{15/2}$	&	$7i_{13/2}$ 	&	154113.2836	&	856.8206	&	8.5185	&	125.2672	&	2.6924	&	\SI{0.314}{\percent}	&		&		\\
	$\bar{p}$ 	&	$^{132}$Xe	&	$10m_{19/2}$	&	$9l_{17/2}$ 	&	170796.0496	&	905.8857	&	8.9370	&	150.2826	&	3.2019	&	\SI{0.353}{\percent}	&		&		\\
	$\bar{p}$ 	&	$^{184}$W	&	$12o_{23/2}$	&	$11n_{21/2}$	&	180885.4799	&	913.8976	&	8.9641	&	166.1718	&	3.5826	&	\SI{0.392}{\percent}	&		&		\\
	$\bar{p}$ 	&	$^{238}$U	&	$14r_{27/2}$	&	$13q_{25/2}$	&	172808.5044	&	810.9335	&	7.8930	&	150.2331	&	3.2202	&	\SI{0.397}{\percent}	&		&		\\
	\end{tabular}
\end{ruledtabular}
\end{table*}			
\endgroup

\begingroup
\squeezetable
\begin{table}							
\caption{Electron screening shift (\si{\electronvolt}) for transitions listed in Table \protect \ref{tab:energies} and relative value with respect to the total QED correction. Screening for $\mu^-$Ne, $\bar p$Ne and $\bar p$Ar, which have no remaining electrons is not shown. For the \num{2}(\num{10})-electron screening, the electrons are  in the  $1s^2$ (resp. $1s^2 2s^2 2p^6$) configurations.}
\label{tab:elec_screen} 							
\begin{ruledtabular}
\begin{tabular}{ccccdrd}	
\multicolumn{1}{c}{Part.}	&	\multicolumn{1}{c}{Elem.}	&	\multicolumn{1}{c}{Initial}	&	\multicolumn{1}{c}{Final}	&	\multicolumn{1}{c}{2 elec. }	&	\multicolumn{1}{c}{\% QED}	&	\multicolumn{1}{c}{10 elec.}	\\
	&		&	\multicolumn{1}{c}{state}	&	\multicolumn{1}{c}{state}	&	\multicolumn{1}{c}{screen.}	&		&	\multicolumn{1}{c}{screen.}	\\
\hline													
	$\bar{p}$ 	&	$^{84}$Kr	&	$13q_{25/2}$	&	$12o_{23/2}$	&	-3.20	&	\SI{-3.47}{\percent}	&	-3.50	\\
 	$\mu^{-}$	&	$^{132}$Xe	&	$6h_{11/2}$	&	$5g_{9/2}$	&	-37.95	&	\SI{-15.26}{\percent}	&	-42.49	\\
	$\bar{p}$ 	&	$^{40}$Ar	&	$6h_{11/2}$	&	$5g_{9/2}$	&	-0.07	&	\SI{-0.01}{\percent}	&	0.18	\\
	$\bar{p}$ 	&	$^{84}$Kr	&	$9l_{17/2}$ 	&	$8k_{15/2}$	&	-1.14	&	\SI{-0.20}{\percent}	&	-1.19	\\
	$\bar{p}$ 	&	$^{132}$Xe	&	$12o_{23/2}$	&	$11n_{21/2}$	&	-5.14	&	\SI{-1.14}{\percent}	&	-5.74	\\
	$\mu^{-}$	&	$^{84}$Kr	&	$4f_{7/2}$	&	$3d_{5/2}$  	&	-6.91	&	\SI{-0.96}{\percent}	&	-7.58	\\
	$\mu^{-}$	&	$^{132}$Xe	&	$5g_{9/2}$  	&	$4f_{7/2}$ 	&	-24.53	&	\SI{-3.66}{\percent}	&	-27.47	\\
	$\bar{p}$ 	&	$^{84}$Kr	&	$8k_{15/2}$	&	$7i_{13/2}$ 	&	-0.80	&	\SI{-0.08}{\percent}	&	-0.79	\\
	$\bar{p}$ 	&	$^{132}$Xe	&	$10m_{19/2}$	&	$9l_{17/2}$ 	&	-3.17	&	\SI{-0.30}{\percent}	&	-3.51	\\
	$\bar{p}$ 	&	$^{184}$W	&	$12o_{23/2}$	&	$11n_{21/2}$	&	-11.07	&	\SI{-1.02}{\percent}	&	-12.69	\\
	$\bar{p}$ 	&	$^{238}$U	&	$14r_{27/2}$	&	$13q_{25/2}$	&	-32.76	&	\SI{-3.40}{\percent}	&	-38.76	\\
	\end{tabular}
\end{ruledtabular}
\end{table}			
\endgroup

Precision studies with exotic atoms are complementary to ongoing BSQED experiments, as these systems are uniquely sensitive to the high-field vacuum polarization, which represents \SI{\approx 90}{\percent} of the BSQED contribution to the transition energy. At the Schwinger limit, \ie for electromagnetic field strengths exceeding \SI{\sim 1.32E18}{\volt\per\meter}, QED becomes non-linear and eventually real electron-positron pair production is expected from the polarization of the vacuum \cite{sch1951}. This phenomenon has yet to be seen experimentally, but is pursued through collisions of HCI \cite{gkzt2017} and with high-intensity ultrafast lasers (see, \eg \cite{mas2006,dmhk2012,kah2016} and Refs. therein). Considering the final level of transitions in the energy range of interest, we see that for the $1s$ level of U$^{91+}$ the mean electric field is \num{1.5} times the Schwinger limit. In the exotic atom levels considered here, the mean field experienced by the particle far exceeds this limit. The $3d$ level of $\mu^-$Ar reaches already a ratio of \num{1.4}, while it is \num{4.7}  for the $5g$ level of $\mu^-$Xe, and \num{5.9} for the $7i$ level in $\mu^-$U. For antiprotonic atoms,  the $6h$ level of  $\bar p$Ne reaches the Schwinger limit.  The $5g$ level of  $\bar p$Ar is at \num{13}, the  $11n$ level of  $\bar p$Xe is at  \num{14.7} and the  $16u$ level of  $\bar p$U is at  \num{16.1}. Considering the transitions in $\bar p$Xe shown in Fig.~\ref{fig:QEDvsFNS}~(a), all states with $n<22$ probe fields above the Schwinger limit, and their spectroscopy gives unique access to extremely high-field phenomena. As in HCI the dominant QED contribution comes from the self-energy, combined precision studies of HCI and exotic systems should allow to disentangle the two fundamental QED effects of self-energy and vacuum polarization. 
 
In the present letter we have shown that measuring transitions between circular Rydberg states in exotic atoms allows to test BSQED much more accurately than in HCI, without uncertainties due to the nuclear effects, which currently limit BSQED tests. One can realize  \numrange{1}{2} orders of magnitude gains in sensitivity with these systems, opening for the first time the possibility of probing second-order QED effects across a broad range of $Z$. Using state-of-the-art microcalorimeters and high-precision, reference-free x-ray \cite{kdgs1982,dkib2003} and $\gamma$-ray \cite{dksh1980,hv2000} measurements for calibration, it is possible to reach a few ppm accuracy and to test the structure of the vacuum for systems in which the average field strength largely exceeds the Schwinger limit.

%\begin{equation}
%\Delta E_{QED2}=\left(\frac{\mu }{m_e}\right)^3\left(\frac{\alpha }{\pi }\right)^2\frac{\left ( Z\alpha  \right )^4}{n^3}F_{QED2}\left ( Z\alpha \right )m_e c^2,
%\end{equation}
%\begin{equation}
%E=\frac{2}{3}\left(\frac{\mu_r}{m_{\mu}}\right)^3\frac{(Z\alpha )^2}{n^3}m_{\mu}c^2 \left(\frac{Z\alpha R}{\lambda}\right)^2\delta_{1,0}
%\end{equation}
%The limit on the corresponding coupling constant will be around \num{9} times better, like the $\bar p$ to $\mu^-$ mass ratio.

%\begin{Fig.}
%\includegraphics[width=\linewidth]{Fig.s/pbarAtoms-TES-det-limits}
%\caption{Limits on a scalar boson coupling parameters $y_{u}^2$ assuming realistic experimental accuracy as in \protect Fig. \ref{fig:muon-boson-X}.}
%\label{fig:pbar-boson-X}
%\end{Fig.}
%
%\begin{Fig.}
%\includegraphics[width=\linewidth]{Fig.s/pionAtoms-TES-det-limits}
%\caption{Limits on a scalar boson coupling parameters $y_{u}^2$ for pionic atoms assuming realistic experimental accuracy as in \protect Fig. \ref{fig:muon-boson-X}.}
%\label{fig:pion-boson-X}
%\end{Fig.}
\begin{acknowledgments}
The MCDFGME code was developed with Jean-Paul Desclaux who passed away on August 28th, 2020.
We acknowledge very helpful discussions with Takuma  Okumura, Dan Swetz and Joel L. Ullom.
We wish to thank the Institute of Physics of CNRS for support and RIKEN for providing a \textit{Young Scientist Fellowship} to N.P. 
P.I. is a member of the Extreme Matter Institute (EMMI).
\end{acknowledgments}

%\bibliographystyle{apsrev}

%\bibliography{refs2020}

\end{document}